\documentstyle[11pt,newpasp,twoside,epsf]{article}
\def\edcomment#1{\iffalse\marginpar{\raggedright\sl#1\/}\else\relax\fi}
\reversemarginpar

\begin{document}

\title{The Orbits of Terrestrial Planets in the Habitable Zones of Known Extrasolar Planetary Systems}

  \author{Barrie W Jones and P. Nick Sleep}
\affil{The Open University, Walton Hall, Milton Keynes MK7 6AA, UK}

\begin{abstract}

We show that terrestrial planets could survive in variously restricted regions of the habitable zones of 47 Ursae Majoris, $\epsilon$ Eridani, and $\rho$ Coronae Borealis, but nowhere in the habitable zones of Gliese 876 and $\upsilon$ Andromedae. The first three systems between them are representative of a large proportion of the 90 or so extrasolar planetary systems discovered by mid-2002, and thus there are many known systems worth searching for terrestrial planets in habitable zones. We reach our conclusions by launching putative Earth-mass planets in various orbits and following their fate with a mixed-variable symplectic integrator.
\end{abstract}

\section{Introduction}

It will be some years before we know whether any of the known extrasolar planetary systems have planets with masses the order of that of the Earth. We have therefore used a computer model to investigate whether such planets could be present, and in particular whether they could remain confined to the habitable zones (HZ) of a representative sample of the known extrasolar planetary systems. If so then it is possible that life is present on any such planets.

The HZ is that range of distances from a star where water at the surface of an Earth-like planet would be in the liquid phase. We have used the intermediate boundaries for the HZ originating with Kasting, Whitmire, \& Reynolds (1993).  Because of simplifications in the climate model, these distances are conservative in that the HZ is likely to be wider. For zero-age main-sequence stars (ZAMS stars) the HZ lies closer to the star the later its spectral type. As the star ages the boundaries move outwards.

To test for confinement to the HZ, we launch putative terrestrial planets, labelled EM, into various orbits in the HZ and use a mixed-variable symplectic integrator (Chambers 1999) to calculate the evolution of the orbit. The integration is halted when a terrestrial planet comes within three Hill radii $(3 R_{\rm H})$ of the giant. This is the distance at which a symplectic integrator becomes inaccurate, and is also the distance at which severe orbital perturbation of the terrestrial planet occurs. Confinement requires that the integration lasts for the full pre-set time, $and$ that the semimajor axis of the terrestrial planet remains within the HZ. The pre-set time depends on the shortest orbital period in the system. The integrator requires a time step shorter than about 5\% of this shortest period, and the pre-set time must keep the CPU time per integration less than 100 or so hours. For systems like $\rho$ Coronae Borealis the orbital period of the giant planet is so short that we use 100 Ma (100 million years) as the standard, and also for Gliese 876 and $\upsilon$ Andromedae. For 47 Ursae Majoris and $\epsilon$ Eridani we use 1000 Ma as the standard.

\section{The exoplanetary systems studied}

We selected five systems that between them represent a large proportion of the 90 or so extrasolar planetary systems known by mid-2002. Figure 1 shows the important characteristics of these systems. All five systems have been seen only by Doppler spectroscopy, which yields $m$ ${\rm sin}(i_{0})$ rather than $m$. For an integration we put in an actual mass $m$ for the giant planet. This is tantamount to setting $i_{\rm 0}$ to some value. We used $i_{\rm 0}$ = $90^{\rm o}$, which gives the minimum value for $m$, and other values that varied from one system to another.

All planets are launched with zero mean anomalies but with various longitudes of the periastra. The outcome can be sensitive to the differences $\Delta\varpi$ between these longitudes, in that some differences result in early close encounters (EM comes within $3 R_{\rm H}$) of the giant planet, whereas other differences result in no close encounters within the pre-set integration time.

\begin{figure}
\plotone{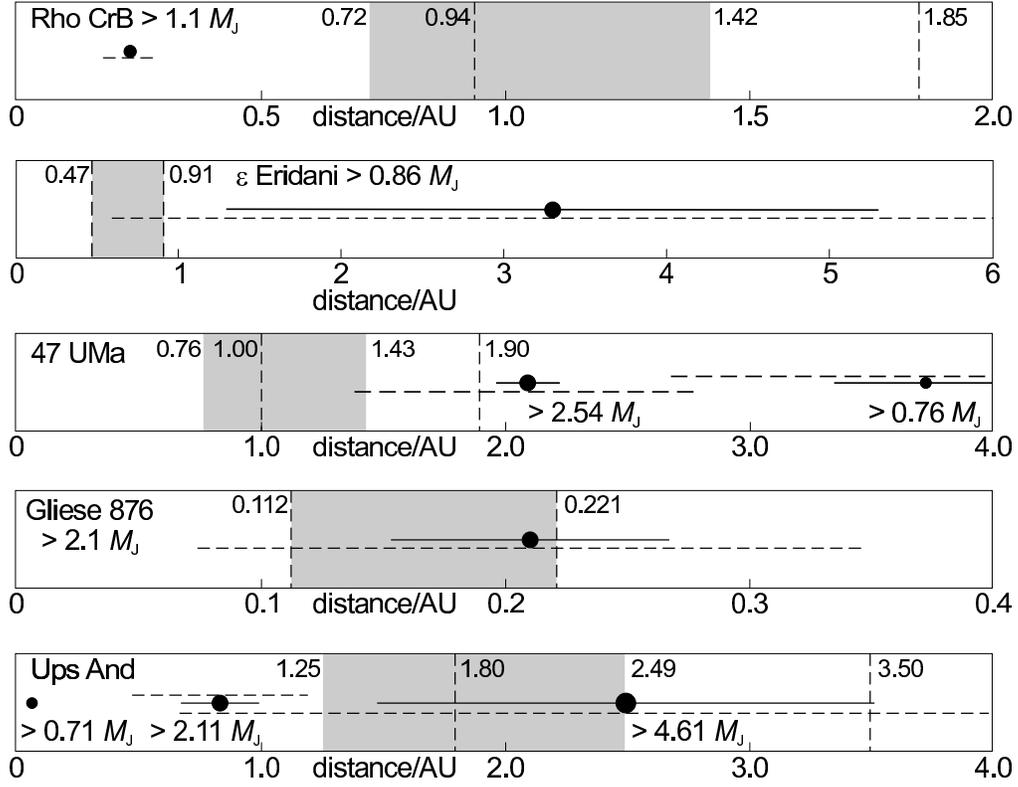}
\caption{Characteristics of the five exoplanetary systems studied. The ZAMS habitable zone HZ(0) is shown shaded, and the boundaries of HZ(now) by vertical dashed lines.  Each giant planet is shown by a black disk labeled with the value of $m$ sin($i_{\rm 0}$), where $m$ is the mass of the giant planet and $i_{\rm 0}$ is the inclination of the planet's orbit with respect to the plane of the sky.  At each giant planet the solid line shows the total excursion $2 \Delta r$  of the giant planet due to its eccentricity.The dashed line extends to $(3 R_{H} + \Delta r)$ each side of the giant planet when it has its minimum mass ($R_{\rm H}$ is proportional to $m^{1/3}$).} 
\end{figure}

\section{Results}

For Gliese 876 and $\upsilon$ And inspection of Figure 1 shows that the whole of HZ(0) and HZ(now) are traversed by $(3 R_{H} + \Delta r)$, and so there should  be no chance of avoiding an early close encounter between EM and a giant planet. This expectation is born out by the integrations, with close encounters no later than about 10 Ma. The second giant now known around Gliese 876 makes things worse.

For $\rho$ CrB at minimum giant planet mass confined orbits for EM are found right across HZ(0), even at the 1:6 mean-motion resonance near the inner boundary of HZ(0). HZ(now) is further out and so here too confinement is ubiquitous. We obtain similar confinement at eight times the minimum giant mass.

The age of $\rho$ CrB is about 6000 Ma, so confinement for the standard pre-set integration time of 100 Ma could be a poor indicator of confinement for 6000 Ma. We therefore extended a few integrations for pre-set times 2-3 times longer than the standard. Full-term confinement was obtained in every case. Moreover, there is no sign of any long-term secular increase in the eccentricity of EM. Therefore, we believe that confinement for 6000 Ma is common, even ubiquitous, in the habitable zones of $\rho$ CrB.

$\epsilon$ Eri is a K2V star and therefore evolves more slowly than the Sun, and it is young, only 500-1000 Ma. Therefore HZ(0) $\simeq$ HZ(now). We explored the inner part of the habitable zone and the region immediately interior to it, including all the mean motion resonances, which are closely spaced. At such resonances there is a strong tendency for $\Delta\varpi_{\rm EMG}$ = $180^\circ$  to give confinement, and for $\Delta\varpi_{\rm EMG}$ = $0^\circ$  to fail to do so, a well-known result.

At minimum giant planet mass, EM orbits that remain confined for the 1000 Ma pre-set time, are restricted to initial ($t$ = 0) semimajor axes of EM, $a_{\rm EM}(0) < 0.59$ AU. Within this distance there is an intricate pattern of outcomes, depending on whether $\Delta \varpi_{\rm EMG}$ = $180^\circ$ or $0^\circ$, and whether $a_{\rm EM}(0)$ is at a mean-motion resonance. Only within 0.44 AU is there general confinement. We also explored 1.39 times the minimum mass. This is because $\epsilon$ Eri has a circumstellar dust ring that is seen to be elliptical. Assuming it to be circular we get an estimate of $i_{\rm 0}$ = $46^\circ$, that corresponds to the 1.39 multiplier. At this greater mass there is general confinement within 0.44 AU, but no confinement at all at $a_{\rm EM}(0) > 0.46$ AU.

47 UMa has an outer giant with a present eccentricity $e < 0.2$. In Figure 1 the value is 0.1. This value was also used in the integrations at $t$ = 0, and also a value of $10^{-5}$. As well as minimum mass, the giant planets were given 1.5 times the minimum, in accord with estimates of $i_{\rm 0}$ for this system (Gonzalez 1998). At 1.5 minimum mass, with a few exceptions, the two giant planets were launched with $\Delta\varpi_{\rm G1G2}$ = $0^\circ$  and $120^\circ$, and $\Delta\varpi_{\rm EMG1}$ = $120^\circ$ and $180^\circ$. This taught us that $\Delta\varpi_{\rm G1G2}$ = $120^\circ$ is less likely to lead to confinement. So, to err on the side of caution in our claims for confinement, at minimum mass we ran configurations other than $120^\circ$ only if there was no confinement at 120\deg.

For both masses there is no confinement beyond 1.25 AU. At minimum mass, at most values of $a_{\rm EM}(0) < 1.25$ AU we obtain confinement at both eccentricities and at all values of the $\Delta\varpi$. At 1.5 times the minimum, with $e(\rm 0)$ = $10^{-5}$ the outcome is similar at $a_{\rm EM}(0) < 1.20$ AU, but at $e(0) = 0.1$ confinement is obtained only at a few values of $a_{\rm EM}(0) < 1.15$ AU and for some $\Delta\varpi$.

47 UMa is about 7000 Ma old (Gonzalez 1998). The pre-set integration time of 1000 Ma is a modest fraction of this, though we see close encounters within 500 Ma in all but a small proportion of cases, and in these cases there are clear upward trends in eccentricity of EM. Therefore, confinement for 1000 Ma is very likely to mean confinement for 7000 Ma.

Further details of our work on all but the $\epsilon$ Eri system are at Jones, Sleep, \& Chambers (2001) and Jones \& Sleep (2002).

\section{Conclusions }

Confinement of a terrestrial planet for the age of the star is likely throughout the ZAMS and present-day habitable zones of $\rho$ CrB. In $\epsilon$ Eri (where the habitable zone has barely moved outwards) only the innermost part of the HZ offers confinement at minimum giant planet mass, and then only for certain launch configurations of the terrestrial body. At 1.39 times this minimum mass, confinement is restricted to the inner boundary of the HZ (and inwards), again for certain configurations only. In 47 UMa confinement is found in the inner part of the present HZ, particularly if either the present eccentricity is much less than 0.1 or if the giant planets' masses are close to the minimum values, or if both conditions hold. Gliese 876 and $\upsilon$ And are devoid of confined orbits in their habitable zones. 

Our conclusions on 47 UMa are in accord with work done by others, though this other work is significantly less extensive (Laughlin, Chambers, \& Fischer 2002; Noble, Musielak, \& Cuntz 2002).

\end{document}